\documentclass[referee]{aa} 
\usepackage{graphicx}
\usepackage{txfonts}
\begin{document}

   \title{Close binary evolution}

   \subtitle{I. The tidally induced shear mixing in rotating binaries}

   \author{H.F. Song,
          \inst{1,2,4}
           A. Maeder\inst{2},
          G. Meynet\inst{2},
          R.Q. Huang\inst{3},
            S. Ekstr\"om\inst{2}
            \and
             A. Granada\inst{2}
          }
\authorrunning{Song et al.}
\institute{College of Science, Guizhou University,
             Guiyang, Guizhou Province, 550025, P.R. China\\
             \email{songhanfeng@163.com;sci.hfsong@gzu.edu.cn}
\and
Geneva Observatory, Geneva University, CH-1290 Sauverny, Switzerland
\and
National Astronomical Observatories/Yunnan Observatory,
     the Chinese Academy of Sciences, Kunming, Yunnan Province, 650011, P.R. China
\and
Key Laboratory for the Structure and Evolution of Celestial Objects, Chinese Academy of Sciences, Kunming, 650011, P.R. China}

   \date{Received; accepted }


  \abstract
   {Tides are known to play a great role in binary evolution, leading in particular to synchronisation
of axial and orbital rotations and to binary mass transfer.}
   {We study how tides in a binary system induce some specific internal shear mixing, able to
substantially modify the evolution of close binaries prior to mass transfer.}
  { We construct numerical models accounting for tidal interactions, meridional circulation, transport of
angular momentum, shears and horizontal turbulence and consider a variety of orbital periods and initial rotation
velocities.}
   {Depending on orbital periods and rotation velocities, tidal effects  may spin down (spin down Case) or spin up
(spin up Case) the axial  rotation.  In both cases, tides may induce a large internal differential rotation. The resulting
tidally induced  shear mixing (TISM)  is so efficient that the internal distributions of angular velocity and chemical
elements are greatly influenced.
The evolutionary tracks are modified, and
in both cases of spin down and spin up, large amounts of nitrogen  can be transported to the stellar surfaces
before any binary mass transfer.
Meridional circulation, when properly treated as an advection, always tends to counteract
the tidal interaction, tending to spin up the surface when it is braked down and vice versa.
As a consequence, {\bf the times needed for the axial angular velocity to become equal to the orbital angular velocity may be larger than given by typical
synchronization timescales.}
Also, due to meridional circulation some differential rotation remains in  tidally locked binary systems.}
   {}


   \maketitle
%

\section{Introduction}
Rotation is  an important
factor to be considered in the evolution
of massive stars (Kippenhahn \& Thomas 1970; Endal \& Sofia 1976;
Meynet \& Maeder 1997; Langer 1998). The centrifugal force not only
gives the star  an oblate shape, but it also induces many
instabilities leading to the mixing of chemical elements  in the stellar
interiors (Maeder \& Meynet 2000; Heger, Langer \&
Woosley 2000a; Huang 2004a,b). In addition to convection and
semi-convection, there are several processes which may contribute to
the mixing of the chemical elements in stellar interiors, such
as shears due to internal differential rotation,
meridional circulation, magnetic instabilities,  magnetic
braking at the surface of the star which favors large shears (Chaboyer \& Zahn 1992; Zahn 1992; Maeder \&
Meynet 2001,2005; Meynet et al. 2011; Mathis et al.
2004a,b). These physical processes are active both in single and
binary stars. They may have important consequences on
observable properties (Langer \& Maeder 1995; Heger \& Langer
2000b; Song et al. 2009, 2011).  They can mix the
material of the core and envelope  leading, among
 other consequences, to nitrogen enrichment at the stellar surface.
Many B- and O-type stars show nitrogen excesses (Walborn 1976; Heap
et al. 2006; Hunter et al. 2009; Przybilla et al. 2010) and the mentioned
instabilities may play a role in these facts.

Recent results suggest that the half of the massive stars in the Tarentula region may
exchange mass with a binary companion, thus potentially largely affecting the course of stellar evolution through tides,
mass transfer and mergers (Sana et al. 2013; de Mink et al. 2013).  The fraction of binary systems,
and  the fraction of short period O-type stars, are largely varying in different clusters (Mahy et al. 2009),
the binary properties being possibly related to the density of the clusters.

Many physical effects studied in the context of single star evolution need also to be considered in the framework
of binary evolution. Here we  concentrate on the interaction of some effects of rotation, such as
meridional circulation, shears, and horizontal turbulence, with binary evolution.
In particular, the tidally induced shear mixing (hereafter called TISM) may be
important in binary evolution. The tidal interactions produce
braking, particularly in the outer stellar layers and thus  may enhance
the internal differential rotation. The instabilities associated to
large shears result in a significant
transport of the chemical elements. In the present work we want to investigate how the tidal  braking
and its  related effects, in particular TISM, can
affect the evolution of close binaries.

On top of that, meridional circulation, when correctly treated
as a circulation and not as a diffusion, often reacts in a interesting way, by even being able to transport
angular momentum from regions  with low rotation  to regions where rotation is fast (Maeder 2009). Thus, by studying TISM
in the framework of  models where meridional circulation is consistently treated, we may see the
reactions of the star to the interaction of meridional circulation and tidal braking. These effects, which were not foreseen,
appear to play a significant role in binary evolution.

The paper is organized as follows: The tidal braking is presented in Section 2.
The equations expressing internal transport of chemical elements and angular momentum are presented in Section 3.
In Section 4, the results of numerical calculation are described and discussed in details.
Finally in Section 5, conclusions are made.

\section{Tidal interactions in rotating binaries}

One may distinguish the equilibrium
and dynamical tides (Zahn 1966; Zahn 1975). For not yet synchronised systems, subject to the effects of dynamical tides,
the dissipation mechanisms play a major role.
These are typically the viscous effect of turbulence  in stars with a
convective envelope and the radiative damping for stars with a radiative envelope (Zahn 1977).
A new expression
for the tidal synchronisation timescale due to a turbulent medium has  been proposed by Toledano et al. (2007),
\begin{equation}
\tau_{\mathrm{sync,turb}}=f_{\rm{turb}}q^{-2}(\frac{R}{a})^{-6}   \mathrm{year},
\end{equation}
where $a$ is the separation between the two components in a binary
system, $f_{\rm{turb}}\sim 1$ depends on the structure of the star.
Equation (1) can be applied to stars with an envelope hosting a
strong turbulence  due to convection and, maybe, due to rotational
instabilities, such as the horizontal turbulence
proposed by Zahn (1992). However, the adequation of this expression to this last case
is still under question, since the viscous effect of horizontal turbulence is about 8 orders of magnitude
smaller than the viscosity of classical convection in massive stars.

The dynamical synchronisation timescale due to radiative damping for
stars with a radiative envelope, {\bf without the contribution of
the equilibrium tide}, has been  given by (Zahn 1977),
\begin{equation}
\tau_{\rm{sync,rad}}=\frac{1}{5\times 2^{5/3}q^{2}(1+q)^{5/6}E_{2}}(\frac{R^{3}}{GM})^{1/2}\beta (\frac{a}{R})^{17/2}.
\end{equation}
Here, $q = M'/M$ is the mass ratio, $M'$ is the mass of the companion
star, $R$ and $M$ are the radius and mass of the primary star considered here, $G$ is the
gravity constant, $\beta=I_{e}/MR^{2}$ is the so-called gyration radius,
{\bf $I_{e}$ is the moment of inertia of
the external layers where tidal energy is dissipated}, and $E_{2}$ is the tidal coefficient,
which is sensitive to the structure of the star, in particular to the size
of the convective core. It can be expressed by (Yoon, Woosley \& Langer 2010)
\begin{equation}
E_{2}\sim 10^{-1.37}(\frac{R_{\rm{conv}}}{R})^{8},
\end{equation}
where $R_{\rm{conv}}$ is the convective-core radius.

A comparison of the above two  dissipation timescales (Eqs 1 and 2), for the
turbulent and radiative cases, has been performed by de Mink et al.
(2009).

\textbf{In the present work, the change of the spin angular momentum due to  tidal
interaction is computed using Eq. (5.6) from Zahn (1977):}
\begin{equation}
\bf{\frac{dI_{e}\Omega}{dt}=-3MR^{2}(\Omega-\omega_{\rm{orb}})(\frac{GM}{R^{3}})^{1/2}[q^{2}(\frac{R}{a})^{6}]E_{2}s_{22}^{5/3}},
\end{equation}
\noindent \textbf{where $I_{e}\Omega $ is the angular momentum of
the external layers where tidal energy is dissipated}, $\Omega$ and
$\omega_{\rm{orb}}$ are respectively  the angular velocity of axial
and orbital rotation. Tidal interactions spin the star down when $
\Omega > \omega_{\rm{orb}}$ and up when $ \Omega < \omega_{\rm
{orb}}$.

Rigorously speaking, the effect of  tidal braking should be applied to the whole radiative envelope.
As some differential rotation is generally present, there may be some deviations from the above current expressions
for  tidal braking. However, some approximations  are justified.
In our current models, we treat a limited fraction of the outer
layers (about $3 \%$ of the total mass ) with the assumptions  of
constant chemical abundances $X_{i}=const.$, constant luminosity
$L=const.$, and angular velocity $\Omega=const.$ These  homogeneous
outer layers of small mass content may encompass  about 30\% of the
total radius, which implies that about $92 \%$ of the total tidal
effect (going like $R^{6}$) is deposited in these layers (for a
value of $20 \%$ of the radius, the deposited fraction would be $79
\%$). This  means that most of the energy dissipated by the tidal
torque is deposited in the outer layers, where the angular velocity
is anyway taken as a constant. So the various approximations made in
the models are internally consistent.

{\bf Assuming $I_{e}$ and $\omega_{\rm orb}$ are constant in Eq. (4),
we can obtain, ${t_{\rm rot}}$, the typical timescale for a decrease by a factor $e$ of the
difference between $\Omega$ and $\omega$, :}
\begin{equation}
-\frac{1}{(\Omega-\omega_{\rm{orb}})}\frac{d (\Omega-\omega_{\rm orb})}{dt}=
\frac{1}{t_{\rm rot}},
\end{equation}
with
\begin{equation}
\frac{1}{t_{\rm rot}}=3 (\frac{GM}{R^{3}})^{1/2}\frac{MR^{2}}{I_{e}}E_{2}[
q^{2}(\frac{R}{a})^{6}]s_{22}^{5/3}.
\end{equation}
\textbf{The synchronization rate $t_{rot}^{-1}$ is a
function of both the strength of the perturbing potential as
measured by the quantity in brackets $[q^{2}(\frac{R}{a})^{6}]$, and
of the tidal frequency
$s_{22}=2(\Omega-\omega_{\rm{orb}})(\frac{R^{3}}{GM})^{1/2}$.}

{\bf In the rest of the paper, we shall call ${t_{\rm rot}}$, the
synchronization timescale (implicitly estimated on the ZAMS), while
the time, given by the present evolutionary models, at which
$\Omega=\omega_{\rm orb}$ will be named the effective
synchronization time. The two times may be quite different since
they do not encompass the same physics: ${t_{\rm rot}}$ only
accounts for the change of $\Omega$ due to the tidal interaction and
is estimated for initial values on the ZAMS, while the
synchronization time accounts for the changes with time of
$\Omega-\omega_{\rm orb}$ and for the evolutions with time of all
the quantities involved in the expressions of ${t_{\rm rot}}$, as
well as for all the other processes (in addition to tidal
interactions) modifying the surface angular velocity of the star
(changes of the radius of the star and all the processes
redistributing the angular momentum inside the star).}

In the case of a circular orbit, the orbital angular momentum of a binary with a orbital separation $a$
is given by
\begin{equation}
J_{\rm{orb}}= M_{1}M_{2}[\frac{Ga}{M_{1}+M_{2}}]^{1/2},
\end{equation}
so that
\begin{equation}
\frac{\dot{a}}{a}=\frac{2\dot{J}_{orb}}{J_{orb}}-2\frac{\dot{M}_{1}}{M_{1}}-2\frac{\dot{M}_{2}}{M_{2}}+\frac{\dot{M}_{1}+\dot{M}_{2}}{M_{1}+M_{2}},
\end{equation}
where a dot indicates time derivation.


\section{Internal transport of chemical elements and of angular momentum}
\subsection{The equation for angular momentum transport}

The transport of angular momentum inside a star is implemented following the prescription of Zahn (1992).
This prescription was complemented by Talon \& Zahn (1997) and Maeder \& Zahn (1998). In the radial direction, it obeys the equation
\begin{equation}
\rho \frac{d}{d t}( r^2 \bar{\Omega})_{M_r} =
   \frac{1}{5r^2} \frac{\partial}{\partial r} ( \rho r^4 \bar{\Omega} U(r) )
   + \frac{1}{r^2} \frac{\partial}{\partial r} ( \rho D r^4 \frac{\partial \bar{\Omega}}{\partial r}).
\end{equation}
The first term on the right hand side of this equation is the divergence of the advected flux of angular momentum, while the second term is the divergence of the diffused flux. $D$ is the total diffusion coefficient in the vertical direction, taking into account the various instabilities that transport angular momentum. The effects of expansion or contraction are automatically
included  in a Lagrangian treatment.
During the evolution, central density increases and the core spin faster, the second term allows, for example, the transport of angular momentum from the core to the surface. It will also transmit the external braking to the interior of the star. The meridional circulation is most efficient for such transports, its velocity was determined by Zahn (1992) and Maeder \& Zahn (1998),
\begin{eqnarray}
U(r)= &\frac{P}{\bar{\rho g} C_P \bar{T}}\frac{1}{[ \nabla_{\rm{ad}} - \nabla_{\rm{rad}} + (\varphi/\delta) \nabla_\mu ]}
[ \frac{L(r)}{M_\star(r)} ( E_\Omega^\star+ E_\mu )+ \frac{C_P}{\delta}\frac{\partial \Theta}{\partial t}],
\end{eqnarray}
where $C_P$ is the specific heat at constant pressure, $\nabla_{ad}=\frac{P \delta}{\rho T C_P}$ is the adiabatic gradient, $M_\star=M (1 - \frac{\Omega^2}{2\pi g \rho_{m}} )$, and $\Theta=\tilde{\rho}/\bar{\rho}$ is the ratio of the variation of the density to the average density on an equipotential. Both $\varphi$ and $\delta$ arise from the  equation of state in the form $\frac{d \rho}{\rho} = \alpha \frac{d P}{P} + \varphi \frac{d \mu}{\mu} - \delta \frac{d T}{T}$, and $E_\Omega$ and $E_\mu$ are terms that depend on the $\Omega$- and $\mu$-distributions respectively. The quantities $E_\Omega^\star$  and $E_{\mu}$ are given by (Maeder 2009)
\begin{eqnarray}
 &E_\Omega^\star  =2[1-\frac{\bar{\Omega^{2}}}{2\pi G\bar{\rho}}-\frac{\bar{\epsilon}+\bar{\epsilon}_{\rm{grav}}}{\varepsilon_{m}}]\frac{\tilde{g}}{\bar{g}}-\frac{\rho_{m}}{\bar{\rho}}\{ \frac{r}{3} \frac{d}{dr}[H_{T}\frac{d}{dr}(\frac{\Theta}{\delta})&\\\nonumber
&-\chi_{T}\Theta +(1-\frac{1}{\delta})\Theta]-\frac{2H_{T}}{r}(\frac{\Theta}{\delta})+\frac{2}{3}\Theta\} & \\\nonumber
&-\frac{\bar{\epsilon}+\bar{\epsilon}_{\rm{grav}}}{\epsilon_{m}}[H_{T}\frac{d}{dr}(\frac{\Theta}{\delta})+(f_{\epsilon}\epsilon_{T}-\chi_{T})(\frac{\Theta}{\delta})]&\\\nonumber  &+(2-f_{\epsilon}-\frac{1}{\delta})\Theta],&
\end{eqnarray}
with
\begin{eqnarray}
 E_{\mu}=\frac{\rho_{m}}{\bar{\rho}}\{\frac{r}{3} \frac{d}{dr}[H_{T}\frac{d}{dr}(\frac{\varphi}{\delta}\Lambda)-(\chi_{\mu}+\frac{\varphi}{\delta}\chi_{T} +\frac{\varphi}{\delta})] \Lambda \\\nonumber
-\frac{2H_{T}}{r} \frac{\varphi}{\delta}\Lambda\}+\frac{\bar{\epsilon}+\bar{\epsilon}_{\rm{grav}}}{\epsilon_{m}}[H_{T}\frac{d}{dr}(\frac{\varphi}{\delta}\Lambda)\\\nonumber
 +(f_{\epsilon}\epsilon_{\mu}+f_{\epsilon}\frac{\varphi}{\delta}\epsilon_{T}-\chi_{\mu}-\frac{\varphi}{\delta}\chi_{T}-\frac{\varphi}{\delta}) \Lambda],
\end{eqnarray}
where the quantities have the same significations as in Maeder \&
Zahn (1998). \textbf{We have taken the same boundary conditions as
Talon et al. (1997; see also Denissenkov et al. 1999; Meynet
\& Maeder 2000)}.

\subsection{The equation for the transport of chemical species}

The horizontal turbulence competes efficiently with the advective term of meridional circulation
for transporting the chemical species (Chaboyer \&  Zahn 1992). The horizontal flow tends to homogenize
the layers in such a way that
the resulting transport of chemical species by both meridional circulation and horizontal turbulence can be computed as a
diffusive process with the coefficient $D_{\rm eff}$. The change of the abundance for a
given chemical element $i$ in the shell with  coordinate $r$ is thus (Zahn 1992):
\begin{equation}
\rho \frac{d X_i}{d t} = \frac{1}{r^2} \frac{\partial}{\partial r}[ \rho r^2 ( D + D_{\rm{eff}}) \frac{\partial X_i}{\partial r}] + ( \frac{d X_i}{d t})_{\rm{nucl}},
\end{equation}
where $D$ is the same as in Eq.(8) and is the total diffusion coefficient in the vertical direction, taking into account the instabilities that transport the chemical elements. The last term accounts for the change in abundances produced by nuclear reactions.
We can then define an effective diffusion coefficient, $D_{\rm eff}$ that combines the effects of  horizontal
diffusion and  of the meridional circulation  (Chaboyer \& Zahn 1992),
\begin{equation}
D_{\rm{eff}} = \frac{1}{30} \frac{| r\ U(r)|^2}{D_{\rm{h}} }.
\end{equation}
We have applied the shellular-rotation hypothesis, which postulates that in differentially rotating stars the angular velocity $\Omega$ is constant on isobars. This results from the strong horizontal turbulence, which  is expected because there is no restoring force in the horizontal (isobaric) direction, as the buoyancy force (the restoring force of the density gradient) acts in the vertical direction. Zahn (1992) relates the diffusion coefficient to the viscosity caused by horizontal turbulence
\begin{equation}
D_{\rm{h}} \approx \nu_{\rm{h}} = \frac{1}{c_{\rm{h}}}\ r\ | 2\,V(r) - \alpha\,U(r)|,
\label{EqDh}
\end{equation}
where $c_{\rm h}$ is a constant of the order of 1, $V(r)$ is the horizontal component of the meridional circulation velocity, $U(r)$ its vertical component (see Eq.9), and in this expression $\alpha = \frac{1}{2} \frac{d \ln (r^2 \bar{\Omega})}{d \ln r}$.

Differential rotation induces shear turbulence at the interface of layers that have different rotational velocities.
A layer remains stable if the excess of kinetic energy due to  differential rotation is inferior to the energy needed to
overcome the stabilizing density gradient in radiative zones
(this is expressed by the \textit{Richardson criterion}). The effects of  thermal dissipation, which reduce  the buoyancy force, are accounted for.
The coefficient of diffusion by shear turbulence was determined by Maeder (1997)
\begin{eqnarray}
D_{\rm{shear}} =f_{\rm{energ}} \frac{H_P}{g\delta} \frac{K}{[\frac{\varphi}{\delta}\nabla_\mu+ ( \nabla_{\rm{ad}} -
\nabla_{\rm{rad}} )]}(\frac{9\pi}{32}\ \Omega\ \frac{{d} \ln \Omega}{{d} \ln r} )^2,
\end{eqnarray}
where $K=\frac{4acT^3}{3\kappa \rho^2 C_P}$ is the thermal diffusivity, and with $f_{\rm energy} \approx 1$, and $\varphi=(\frac{dln \rho}{dln \mu})_{P,T}$.

\section{Physical ingredients of the models}

The treatment of rotation was developed in a series of papers published previously by the Geneva
group (Maeder 1999, 1997; Maeder \& Meynet 2000; Maeder \& Zahn 1998; Meynet \&
Maeder 1997, 2000; see also the recent review by Maeder \& Meynet 2012).

The initial abundances of H, He, and metals are set to X=0.720, Y=0.266, and Z= 0.014.
The mixture of heavy elements is  that by Asplund et al. (2005)
except for the Ne abundance, which is taken from Cunha et al. (2006). The nuclear reaction rates
are generated with the NetGen tool \footnote{http: //www-astro.ulb.ac.be/Netgen/}. They originate
mainly from the Nacre database (Angulo et al. 1999), although some have been redetermined more
recently and updated (see Ekstr\''om et al. 2012 for more details).
The convective zones are determined with the Schwarzschild criterion.
For the H- and He-burning phases, the convective core is extended by overshooting over a distance
$d_{\rm over} =  0. 10 H_{\rm p}$, with $H_{\rm p}$ the local pressure scale height.

The evolution of a binary system consisting of a 15  and a 10 M$_\odot$  star is investigated in
this paper. We assume that the orbit is circular and that the spin axis of the model stars is perpendicular
to the orbital plane. We focus on the evolution of the primary star.
From Eq. (7), assuming that the mass loss by stellar winds is negligible before any mass transfer, one has
for the cases considered here that
\begin{equation}
\frac{\dot{a}}{a}=\frac{2\dot{J}_{\rm{orb}}}{J_{\rm{orb}}}\simeq {10^{-2}-10^{-3} \over 10^{7} \; {\mathrm{years}}},
\end{equation}
therefore the variation of the orbital
separation between the two components is negligible and is assumed to
keep a constant value during the evolution of the binary system.

The evolution is followed from the
onset of central H burning to the moment of Roche Lobe Overflow (RLOF).
Starting from an initial solid body rotation, we then calculate
the evolution of the angular velocity inside the model
consistently accounting for the various transport mechanisms  presented in Sect.~3 and for the
tidal interactions between the two stars.

Unlike stellar winds, the tidal braking does not only spin down the star, but it may also spin it up. We have modeled the tidal mixing in two series of evolutionary sequences
\begin{itemize}
\item {\bf Case spin down:} The orbital periods and the initial rotation velocities have been chosen
so that tidal interaction spin down the primary.
The initial velocity is chosen as $\upsilon_{\rm ini}/ \upsilon_{\rm crit}$=  0.6,  $\upsilon_{\rm ini}$  is the equatorial surface velocity
on the ZAMS and $\upsilon_{\rm crit}$  is the critical velocity at the same stage (the critical velocity
is defined as the equatorial velocity such that the centrifugal acceleration exactly balances the
gravity at equator). The consequences of tidal braking were explored for different initial orbital periods ranging
from 1.1 to 1.8 days. For comparison, a
model starting with $\upsilon_{\rm ini}/ \upsilon_{\rm crit}$=  0.6 on the ZAMS,
with no account of tidal braking,  has also been computed.
\item {\bf Case spin up:} The orbital periods and the initial rotation velocities have been chosen
so that tidal interaction spin up the primary.
The initial velocity is taken as $\upsilon_{\rm ini}/ \upsilon_{\rm crit}$ =  0.2.
The initial orbital periods were chosen between 0. 9 and 1. 4 days. A model
without tidal interaction with $\upsilon_{\rm ini}/ \upsilon_{\rm crit}$ =  0. 2  has also been computed.
\end{itemize}

\section{Case spin down : tidal braking}

\subsection{Impact on rotation}

\begin{figure}
   \centering
    \includegraphics[width=8.0cm]{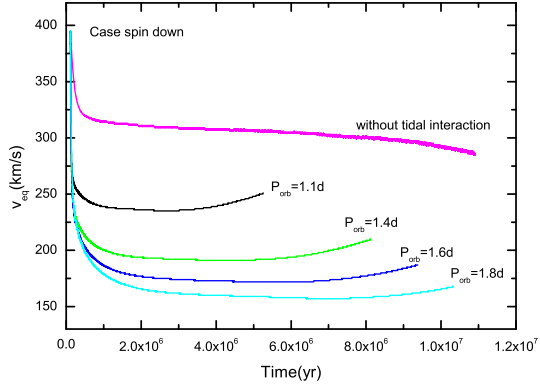}
      \caption{Evolution as a function of time of the surface equatorial velocity for a 15 M$_\odot$ star with
      $\upsilon_{\rm ini}$=0.6 $\upsilon_{\rm crit}$ and a 10 M$_\odot$ companion for different initial orbital periods.
      The case without any tidal interaction  is also shown.}
         \label{veq1}
   \end{figure}

   \begin{figure}
   \centering
    \includegraphics[width=8.5cm]{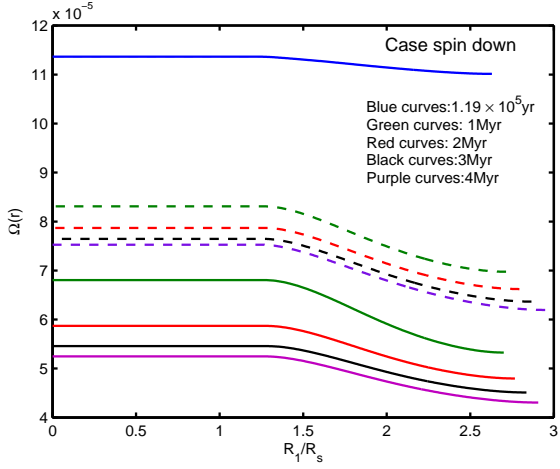}
      \caption{Variation of the angular velocity as a function of the radius in solar units.
      The continuous lines refer to the spin down case with an orbital period equal to 1.8 days.
      The dashed curves correspond to the spin down case with an orbital period equal to 1.1 days.
      Different ages are plotted with different colors as indicated in the inset,
      for both the continuous and dashed lines. The curve corresponding to an age of 119 000 years is the same for the model
      in the system with an orbital period $P$ of 1.1 days and in the one with an orbital period of 1.8 days. }
         \label{omega1}
   \end{figure}

The time evolution of the surface velocity is plotted for various
orbital periods in Fig.~\ref{veq1}. {\bf We see the rapid decrease
of the velocities due to tidal braking.} Synchronization is realised
near the minimum of the curves. After that point, the angular
velocity $\Omega$ at the surface is maintained approximately at the
value imposed by the orbital angular velocity and thus remains
constant. Since, during the Main-Sequence phase, the radii of the
star progressively increase, the surface velocities also slightly
increase after synchronisation. From Fig.~\ref{veq1}, we also see
that:
\begin {itemize}
\item The surface velocity of tidally braked stars is much lower, at a given age, than the surface velocity of an isolated star starting its evolution on the ZAMS
with the same initial velocity. This illustrates the well known fact that, for these orbital periods, the tidal interactions are very strong.
\item Synchronisation timescales are shorter in orbital systems with shorter periods, because  tidal torques are stronger.
\item At the time of synchronisation, lower surface velocities are evidently obtained in systems with longer orbital periods.
\end{itemize}

The evolution of the angular velocity inside models of different ages and in systems with various orbital periods is shown in Fig.~\ref{omega1} up to $4 \cdot 10^6$ yr.
We can see in the model with an orbital period, $P_{orb}=$ 1.1 days,
how tidal braking imposes, after about 1 million years, an angular velocity at the surface converging around $6.6 \times 10^{-5}$ s ($\Omega=\frac{2\pi}{P}=6.6 \times 10 ^{-5}$ s).
The same kind of convergence occurs in the various models but at later times.
{\bf Clearly, the angular velocity is decreased everywhere in the interior of the star by tidal braking, not only at the surface (compare the continuous and dashed curves in Fig. 2
which show the differences due to different tidal torques; when the braking is stronger, the core is also more strongly slowed down). This comes from the coupling
due mainly to meridional currents.}
Most interestingly, some gradient of $\Omega$ is maintained in the star at the time of synchronisation, as a result
of meridional circulation as discussed below.

\begin{figure}
   \centering
    \includegraphics[width=7.8cm]{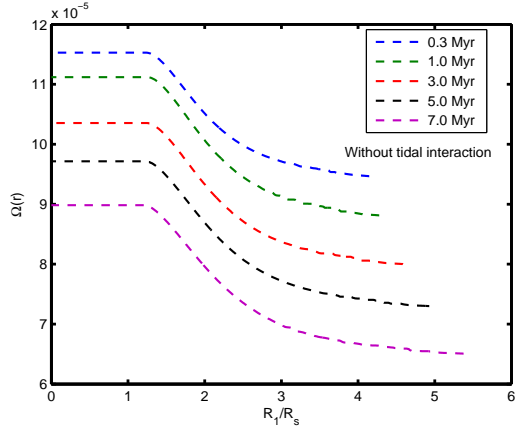}
      \caption{Variation of the angular velocity ([s$^{-1}$]) as a function of the radius and age inside a 15 M$_\odot$ model
      with $\upsilon_{\rm ini}/ \upsilon_{\rm crit}$=  0.6 computed without tidal interaction. }
         \label{fig2a}
   \end{figure}

\begin{figure}
   \centering
    \includegraphics[width=7.8cm]{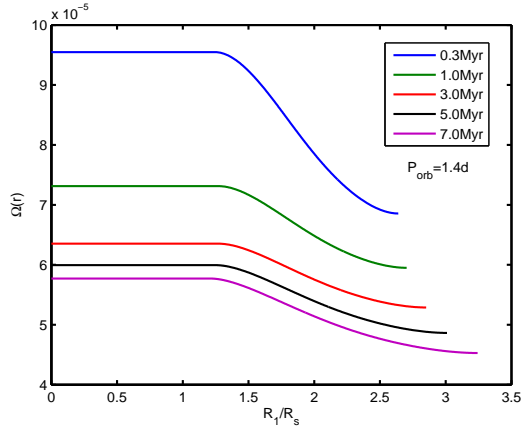}
      \caption{Same as Fig.~\ref{fig2a} for a 15 M$_\odot$ model with tidal braking. The companion is a 10 M$_\odot$ star
      and the orbital period is 1.4 days. Note that the maximum radius is not coinciding with the total radius of the star, only
      a portion of the interior is represented.}
         \label{fig2b}
   \end{figure}

We may also compare the gradients of $\Omega$ in the models with and
without  tidal interactions (see Figs.~\ref{fig2a} and \ref{fig2b}). First, we note  from Fig.~\ref{fig2b} that at the  beginning of the braking, the gradients of $\Omega$ are  stronger in the tidally braked models. This will
imply  stronger mixing of the chemical elements before synchronisation.
Second, we see that, after synchronisation (after about $3.0\cdot
10^6$ years), the gradients of $\Omega$ are shallower in the tidal
braked models.

This last situation results from the interplay of many physical processes.
Indeed, there is an impressive number of effects influencing $\Omega$
inside the models: the structural changes, the convective transport, the shear turbulence which tends to erode the
$\Omega$--gradients, the meridional circulation which can smooth or build up the $\Omega$--gradients, the tidal effects which can remove or bring angular momentum.
Mass loss can also remove angular momentum, especially when a magnetic field
is present in outer layers. One of these effects may be dominant at a given moment at some point inside the star, while part of these effects or all  can
interact in other periods/regions of the star, so it is very difficult to identify a clear cut cause for the shallower $\Omega$-gradients in post synchronised systems.

\begin{figure}
   \centering
    \includegraphics[width=9.0cm]{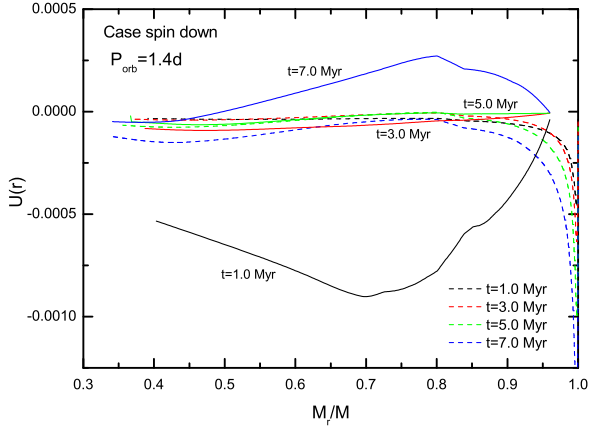}
      \caption{Amplitude of the vertical component of the meridional circulation velocity in cm per second in a 15 M$_\odot$ model at various stages. The dashed lines correspond to a rotating model with $\upsilon_{\rm ini}$=0.6 $\upsilon_{\rm crit}$ and computed with no account of tidal interaction. The continuous lines show the situation when tidal interactions are accounted for, assuming a close companion of 10 M$_\odot$ orbiting the primary with an initial period of 1.4 days.
      Different ages are plotted with different colors as indicated in the inset, for both the continuous and dashed lines.
      The different colors have the same significations as in Fig.2. }
         \label{u1}
   \end{figure}

Let us examine carefully the effects and behavior  of the meridional
circulation in relation with tidal effects. In Fig.~\ref{u1}, the
variation as a function of the Lagrangian mass of $U_2$
\footnote{$U_2(r)$ is defined by $U(r, \theta)=U_2(r) P_2(\cos
\theta)$, where $U$ is the vertical component of the velocity of the
meridional current, $P_2$, the second Legendre polynomial and
$\theta$, the colatitude.} is shown for the spin down Case with an
initial orbital period of 1.4 days (look at the continuous lines). A
negative value of $U_2(r)$  corresponds to a net transport of
angular momentum from the core to the envelope. A positive value
corresponds to the reverse situation, i.e. to  a net transport of
angular momentum from the envelope to the core. We see that at the
beginning of evolution (curves for the ages 1 and $3 \cdot 10^6$
years in Fig.~\ref{u1}), the meridional circulation transports
angular momentum from the core to the envelope. Thus, while tidal
braking slows down the envelope, circulation tends to counteract
this effect by accelerating it. After about $5 \cdot 10^6$ years,
meridional circulation changes sign and the  transport of angular
momentum is from the envelope to the core, favoring the increase of
the ratio between $\Omega$ in the core and that in the envelope, at
a time where synchronisation  tends to homogenize the internal
rotation


Comparing the variation of $U_2$ in the models with and without tidal interaction, one sees important differences.
In the model without tidal
interaction, for the range of ages considered, $U_2$ is always very negative, thus there is a a net transport of angular momentum from the rapid spinning core
to the slow rotating envelope. Such a process goes in the same direction as would do a diffusive process.
In the model with tidal interactions, in a first period until $\sim 5 \cdot 10^6$ years, meridional circulation
transports angular momentum from the core to the envelope, as in the model without tidal
interaction, but with much larger negative  velocities,
as can be seen in Fig.~\ref{u1}. Then, as the circulation changes sign, the transport of angular momentum goes
from the envelope to the core. This last process can no long be modeled through a diffusive process. Thus,
it is essential to describe circulation currents as an advective process.

This redistribution of the angular momentum inside the star by
meridional circulation has consequences for the synchronisation
time. Since circulation tends to counteract the braking of the
surface by the tidal torque, it increases the time needed for
achieving synchronisation. {\bf In Table~\ref{sync1}, two times are
indicated. In the second column  ${t_{\rm rot}}$ is given for the
various initial orbital periods and for ZAMS values. The third
column shows the effective time at which $\Omega$=$\omega_{\rm orb}$
as given by the present evolutionary stellar models.}

{\bf The values of  ${t_{\rm rot}}$ are extremely short, much less than 1\% of the Main-Sequence lifetime of a 15 M$_\odot$ star.
This is quite consistent with the very small timescales obtained by  de Mink et al.
(2009) and also with the very rapid decrease of the surface velocity shown in our present models (see Fig. 1).}

{\bf The effective time at which $\Omega$=$\omega_{\rm orb}$ is much
longer than ${t_{\rm rot}}$ by large factors greater than 70. The
main reason making this time longer is the  outward transport of
angular momentum by circulation, which counteracts  tidal braking.
It must be noted however that looking at Fig.~1, the decrease of the
equatorial surface velocity is quite fast and from an observational
point of view the star may be considered as synchronized well before
the time at which one has the strict equality $\Omega$=$\omega_{\rm
orb}$. Thus, the present results do not contradict the observed fact
that all binaries of the sample of B-type stars observed by Abt et
al. (2002) with periods less than 2.4 days (18 stars) are
synchronized.}


The main consequences of these findings are :
\begin{itemize}
\item Tidal braking has a large impact on meridional currents, which in turn
redistributes the angular momentum inside the star.
\item {\bf ${t_{\rm rot}}$ as defined above greatly underestimates the time needed for $\Omega$ to become equal to $\omega_{\rm orb}$.}
\item The gradient of $\Omega$ is not flat when synchronisation is achieved.
\end{itemize}

\begin{table}[htdp]
\caption{\bf Synchronisation timescales, $t_{\rm rot}$, as given by
Eq. (6) for ZAMS values and time at which $\Omega$=$\omega_{\rm
orb}$ from the numerical models for 15 M$_\odot$ starting with an
initial equatorial velocity on the ZAMS equal to 390 km $s^{-1}$.}
\begin{center}
\begin{tabular}{|c|c|c|}
\hline
P        & $t_{\rm rot}$ (yr) & $t(\Omega=\omega$) (yr)  \\
\hline
      &            &          \\
1.1 & 27 400     & 1 900 000 \\
1.4 & 38 000     & 3 440 000 \\
1.6 & 54 900     &  4 580 000 \\
1.8 & 77 800     & 5 400 000 \\

\hline
\end{tabular}
\end{center}
\label{sync1}
\end{table}%

\begin{figure}
   \centering
    \includegraphics[width=7.8cm]{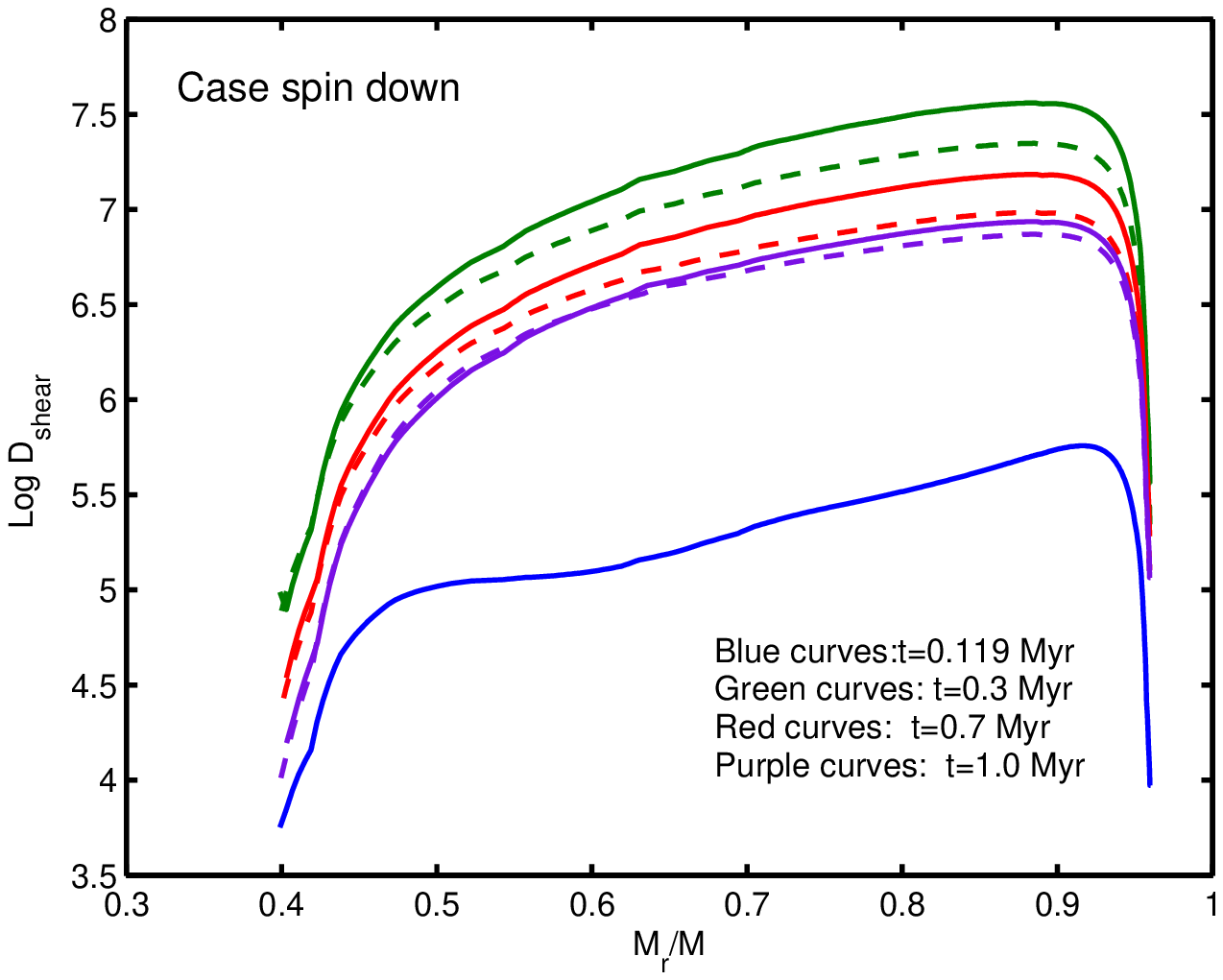}
      \caption{Variation as a function of the Lagrangian mass coordinate of the shear diffusion coefficient in the radiative envelope
      of our 15 M$_\odot$ model. The continuous lines refer to the spin down case with an initial orbital period of 1.8 days.
      The dashed lines show the case when the initial orbital period is equal to 1.1 days. The curves for the periods equal to 1.1
      and 1.8 days, corresponding to an age of 119 000 years, are the same and thus are superposed.
      The different colors has the same significations as in Fig. 2.}
         \label{dshear1}
   \end{figure}

\begin{figure}
   \centering
    \includegraphics[width=7.8cm]{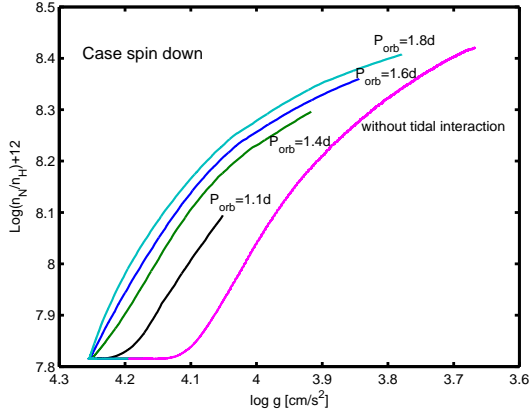}
      \caption{Variation as a function of time of the ratio in number, at the surface, of nitrogen to hydrogen for 15 M$_\odot$ stellar models with (spin down case) and without tidal interactions. The curves are labeled with the values of the initial orbital period.}
         \label{gn1}
   \end{figure}

\subsection{Impact on the chemical composition and evolutionary tracks}

Vertical shear turbulence is the main driver for chemical mixing.
Let us see how  the shear diffusion coefficient varies as a function
of the Lagrangian mass coordinate in models with different tidal
interactions (Fig.~\ref{dshear1}). At the beginning, the diffusion
is quite small since the $\Omega$--gradients are very small. Then
the braking produces a strong shear and the diffusion increases. One
notes that after about 0.2 Myr, the shear diffusion coefficient is
slightly larger in the model with the longer orbital period,
indicating that stronger $\Omega$--gradients  are produced by tidal
interactions in these systems. This probably comes from the fact
that torques in longer period systems are active over longer
durations than in shorter period systems. 

As a result of the $\Omega$-gradients induced by the tidal interactions, the surface becomes enriched in nitrogen
(see Fig.~\ref{gn1}).
In stars with tidal braking a given enrichment is reached for larger values of the surface gravity than in single star starting with the same initial rotation,
indicating a much more efficient mixing.
Interestingly enough, these models produce  relatively  high nitrogen abundances for high log $g$ values, but low surface velocities (see Fig.~\ref{veq1}).
Thus, tidal mixing strongly changes  the correlation between surface velocity, nitrogen abundance and surface gravity with respect to the correlations obtained in models
without tidal mixing.


\begin{figure}
   \centering
    \includegraphics[width=8.5cm]{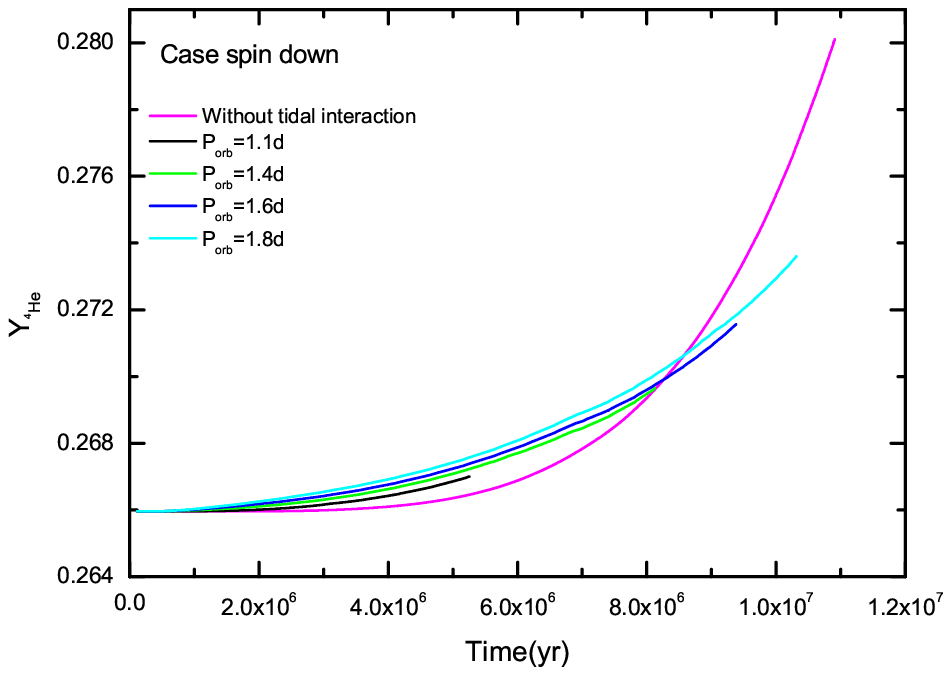}
      \caption{Variation as a function of time of the surface helium mass fraction for 15 M$_\odot$ with (spin down case)  and without tidal interactions. The curves are labeled with the values of the initial orbital period.}
         \label{he1}
   \end{figure}

\begin{figure}
   \centering
    \includegraphics[width=8.5cm]{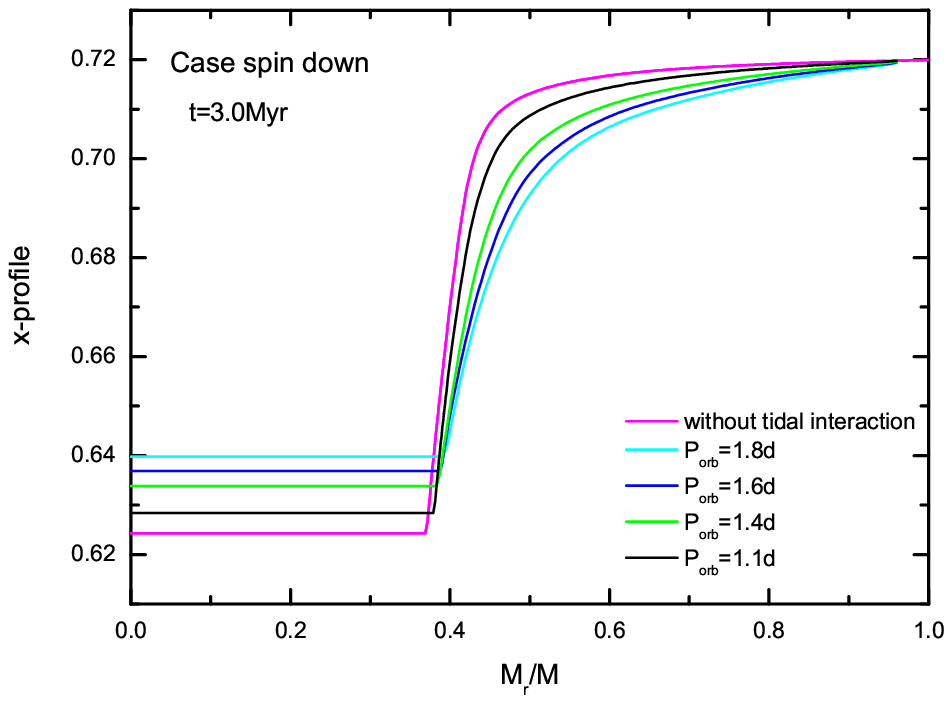}
    \caption{Variation as a function of the Lagrangian mass coordinate of the mass fraction of hydrogen in 15 M$_\odot$ models with (spin down case) and without tidal interaction at the same age during the Main-Sequence phase. Curves for different initial orbital periods are shown.}
         \label{xh1}
   \end{figure}

\begin{figure}
   \centering
    \includegraphics[width=8.5cm]{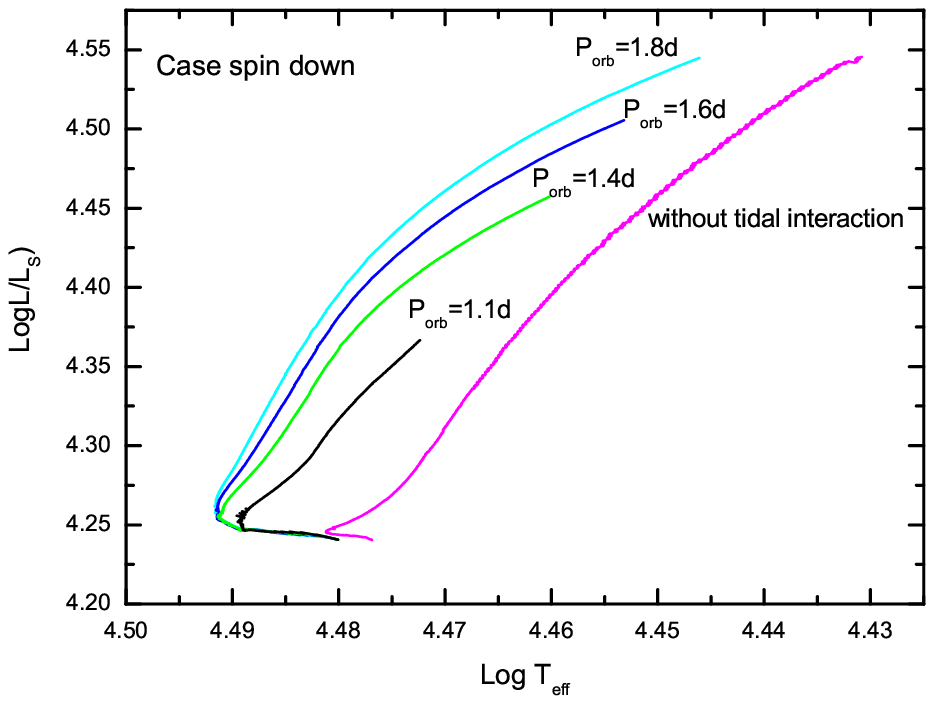}
      \caption{Evolutionary tracks for 15 M$_\odot$ models with (spin down case) and without tidal interaction. Curves for different initial orbital periods are shown.
      For close binaries, only the part of the track corresponding to the period before the Roche Lobe Overflow is plotted.}
         \label{hr1}
   \end{figure}

In Fig.~\ref{he1}, the analog of Fig.~\ref{gn1} is shown but for the surface enrichments in helium. One sees the same qualitative behaviors as for nitrogen, but with smaller  amplitudes, likely not observable.

The different mixings undergone by the stars produce internal compositions which greatly differ at a given age. This can be seen in Fig.~\ref{xh1} where the variations of the mass fraction of hydrogen in our 15 M$_\odot$ are shown at a given age for various efficiencies of the tidal braking. We see that in short period systems, the hydrogen profiles is less affected than in longer period systems. This directly results from the point raised above indicating that
shear mixing is more efficient in longer period systems.


The consequences of these various hydrogen compositions on the evolutionary tracks in the HR diagram can be seen in Fig.~\ref{hr1}. {\bf All tracks with tidal interactions
are shifted to the blue with respect to the track without tidal interaction. The longer the initial period, the greater the blueshift is, consistent with larger mixing in longer period systems.}
The mass-age-luminosity relation is significantly influenced by tidal mixing, in the sense of an overluminosity for a given mass in the case of tidal braking.

\section{Case spin up : tidal acceleration}

\subsection{Impact on rotation}

\begin{figure}
   \centering
    \includegraphics[width=8.5cm]{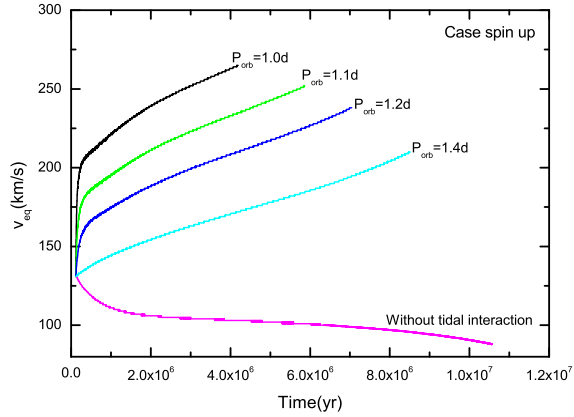}
      \caption{Evolution as a function of time of the surface equatorial velocity for a 15 M$_\odot$ star with $\upsilon_{\rm ini}$=0.2 $\upsilon_{\rm crit}$ and a 10 M$_\odot$ companion for different initial orbital periods. The case without any tidal interaction  is also shown.}
         \label{veq2}
   \end{figure}

\begin{figure}
   \centering
    \includegraphics[width=8.5cm]{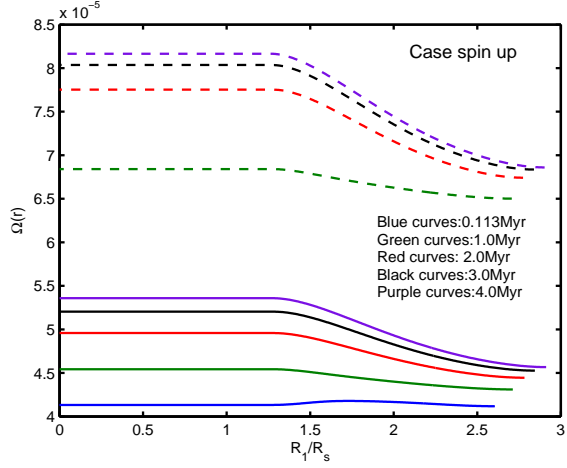}
      \caption{Variation of the angular velocity as a function of the radius in solar units. The continuous lines refer the spin up case with an orbital period equal to 1.4 days. The dashed curves correspond to the spin up case with an orbital period equal to 1.0 days. Different ages are plotted with different colors as indicated in the inset, for both the continuous and dashed lines. The curves for the periods equal to 1.0 and 1.4 days, corresponding to an age of 113 000 years, are the same and thus are superposed. }
         \label{omega2}
   \end{figure}

\begin{figure}
   \centering
    \includegraphics[width=9.0cm]{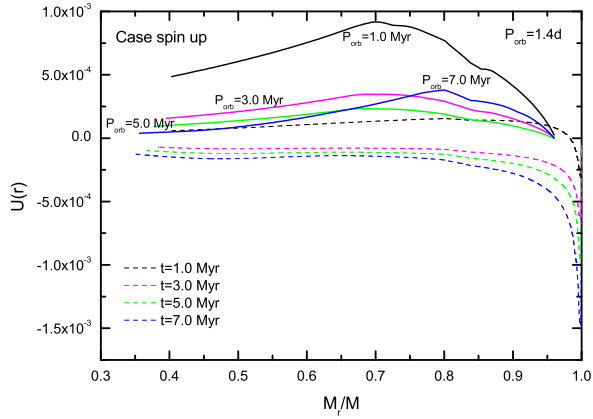}
      \caption{Same as Fig. 1 for a 15 M$_\odot$ model with $\upsilon_{\rm ini}$=0.2 $\upsilon_{\rm crit}$. The continuous lines show the situation when tidal interactions are accounted for assuming a close companion of 10 M$_\odot$ orbiting the primary with an initial period of 1.4 days. The dashed lines correspond to a rotating model computed with no account of tidal interaction. The different colors have the same significations as in Fig.2.}
         \label{u2}
   \end{figure}

Now we discuss situations when the primary is spun up by tidal interactions.
Figure~\ref{veq2} shows how the surface velocity varies as a function of time in case of tidal spin up.
The models with tidal interactions rapidly become fast rotators under the influence of their companions.
As expected, when the orbital period is short,  the surface velocity increase very rapidly and reaches
a high value.

The evolution of $\Omega$ inside different tidally spin up models
are shown in Fig.~\ref{omega2}. {\bf We can see how tidal interaction
imposes the surface angular velocity to
approach the value of
the orbital velocity which is $7.26
\cdot 10^{-6}$ s$^{-1}$ for the system with an orbital
period of 1.0
days, and $5.19\cdot 10^{-6}$ s$^{-1}$, for the system with
an orbital period of 1.4 days. Actually, one sees that these systems
reach Roche Lobe Overflow before being synchronized.}
We see that the
gradient of $\Omega$ increases slightly when time goes on.

In Fig.~\ref{u2},
the radial component  of the vertical velocity of meridional circulation is shown. During the whole period considered,
circulation transports angular momentum from the envelope to the core.
Thus, we obtain in case of tidal spin up, the same kind of general behavior we obtained in case of tidal spin down: meridional circulation
counteracts the effect of the tidal interaction. When the surface is spun down, circulation tends to accelerate it, and when
the surface is spun up, it tends to slow it down. Thus, for both tidal spin up and down,  the synchronisation times
are increased due to the redistribution of the angular momentum inside the star by meridional circulation.

{\bf From Table~\ref{sync2}, we see that the synchronization
timescales, $t_{\rm rot}$, is much shorter than the time at which
the Roche Lobe Overflow (RLOF)  occurs. On the other hand, the RLOF
occurs before the effective time when $\Omega=\omega_{\rm orb}$.
That means that the values shown in the third column of Table~2 are
lower limit for the times $t(\Omega=\omega_{\rm orb})$. Thus we have
a very similar situation than in the case of spin down, where
$t_{\rm rot}$ is much smaller than the time at which
$\Omega=\omega_{\rm orb}$. In contrast however with the case of spin
down, here we cannot say that very rapidly $\Omega$ converges
towards values near $\omega_{\rm orb}$. It does therefore appear
more difficult to reach synchronization in tidally spin up systems.
The reason is likely the following ones:
An inwards transport of a given mass carries more angular momentum than an outwards transport
of the same mass, because
the specific angular momentum increases as a function of the distance to the center.
Thus, it is easier for the circulation  to slow down the surface when it is accelerated by the tidal interaction than for the opposite situation in the case of spin down.}

\begin{table}[htdp]
\caption{\bf Synchronisation timescales, $t_{\rm rot}$, as given by
Eq. (6) for ZAMS values and time at which a Roche Lobe Overflow
occurs for 15 M$_\odot$ starting with an initial equatorial velocity
on the ZAMS equal to 130 km $ s^{-1}$.}
\begin{center}
\begin{tabular}{|c|c|c|}
\hline
P        & $t_{\rm rot}$ (yr) & $t$(RLOF) (yr)  \\
\hline
         &                              &             \\
1.0 &   26 400 &    4 167 000 \\
1.1 &   57 200 &    5 856 000 \\
1.2 &  122 700 &   7 029 000 \\
1.4 &   608 400&   8 496 000 \\
\hline
\end{tabular}
\end{center}
\label{sync2}
\end{table}%

In this case of tidal spin-up, we also see  that a proper advective treatment of meridional circulation is needed for the whole period.
A diffusive treatment   would  not account for the direction of the transport of angular momentum at any moment and thus would even produce effects with the wrong sign.
Incidentally,  Fig.~\ref{u2} also shows that the velocity of the circulation  currents  tend to decrease with time. This reflects the fact that
the gradient of $\Omega$ increases slightly when time goes on (see Fig.~\ref{omega2}).



\begin{figure}
   \centering
    \includegraphics[width=8.0cm]{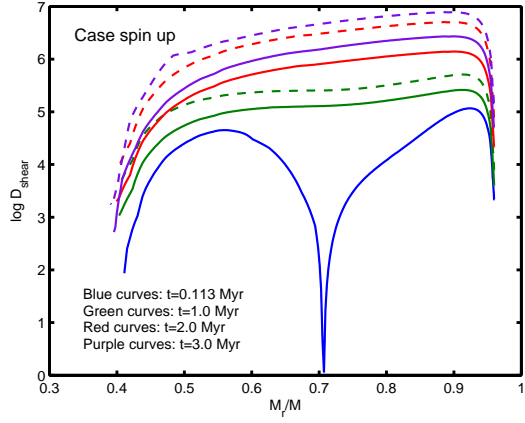}
      \caption{Variation as a function of the Lagrangian mass coordinate of the shear diffusion coefficient in the radiative envelope of our 15 M$_\odot$ model. The continuous lines refer to the spin up case with an initial orbital period of 1.4 days. The dashed lines show the case when the initial orbital period is equal to 1.0 days. The curves for the periods equal to 1.0 and 1.4 days, corresponding to an age of 113 000 years, are the same and thus are superposed. The different colors have the same significations as in Fig.2.}
         \label{dshear2}
   \end{figure}

\begin{figure}
   \centering
    \includegraphics[width=7.8cm]{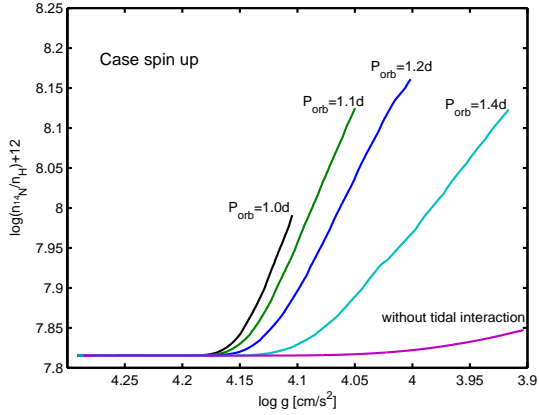}
      \caption{Variation as a function of time of the ratio, in number,  of nitrogen to hydrogen at the surface of a 15 M$_\odot$ stellar model with (spin up case)  and without tidal interactions. The curves are labeled with the values of the initial orbital period.}
         \label{gn2}
   \end{figure}

\subsection{Impact on chemical composition}

Tidal acceleration boosts the shear diffusion  and the surface nitrogen enrichments as can be seen in Figs.~\ref{dshear2} and \ref{gn2}.
From Fig.~\ref{dshear2}, one sees that the shear diffusion coefficient is larger in shorter period systems. This is the reverse of what is obtained
in systems which are tidally spun down. In ``spin down'' system, we had larger $D_{\rm shear}$ in longer period systems as a result of
a smaller, but longer active torque acting on the surface. In ``spin up'' systems, we have larger  $D_{\rm shear}$ in shorter period systems as a result of
a shorter active, but stronger torque acting on the surface.

\begin{figure}
   \centering
    \includegraphics[width=8.5cm]{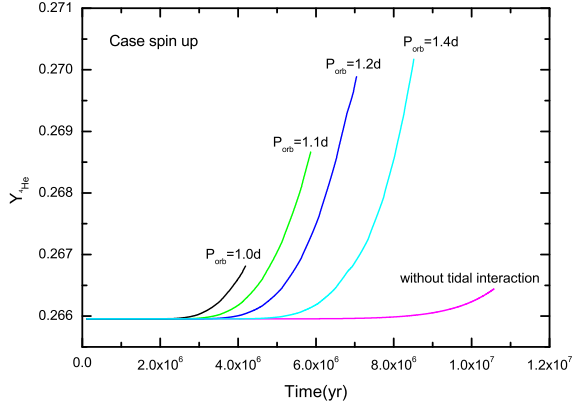}
      \caption{Variation as a function of time of the surface helium mass fraction for 15 M$_\odot$ with (spin up case)  and without tidal interactions. The curves are labeled with the values of the initial orbital period.}
         \label{he2}
   \end{figure}

\begin{figure}
   \centering
    \includegraphics[width=9.0cm]{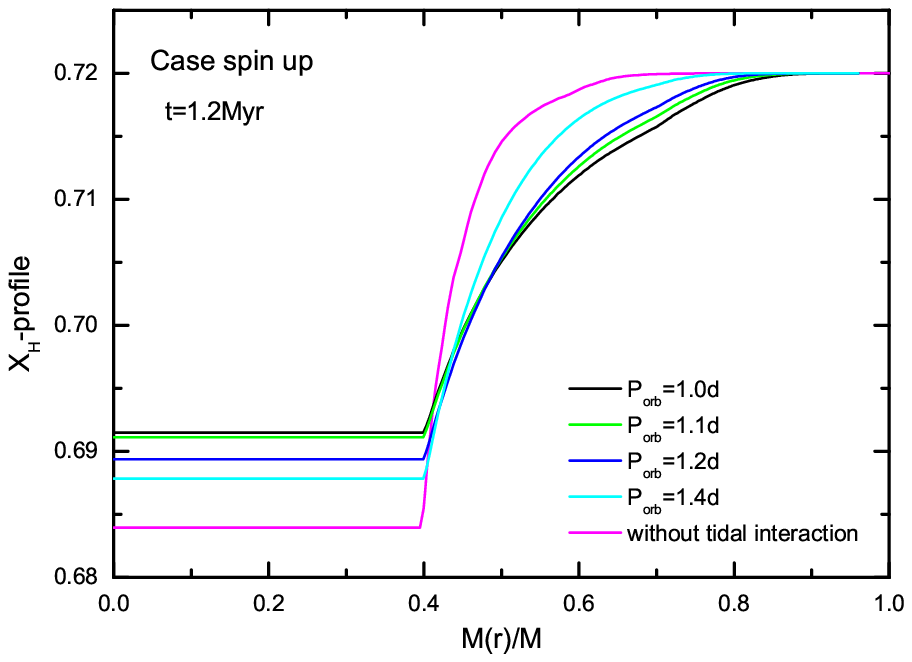}
      \caption{Variation as a function of the lagrangian mass coordinate of the mass fraction of hydrogen in 15 M$_\odot$ models with (spin up case) and without tidal interaction at the same age during the Main-Sequence phase. Curves for different initial orbital periods are shown.}
         \label{xh2}
   \end{figure}

As a consequence of the shears in tidally interacting systems, strong surface enrichments in nitrogen are resulting, as can be seen in Fig.~\ref{gn2}.
A given nitrogen enhancement is obtained earlier, i.e. for lower surface gravity values in shorter period systems.
The surface helium enrichments reflect the same trend, but with limited amplitudes  (see Fig.~\ref{he2}).
The variations of the mass fraction of hydrogen are shown in Fig.~\ref{xh2}.
Consistently with other results, we see that for smaller  initial periods, the effects of the tidal spinning up are stronger, again due the higher $\Omega$--gradients.

\begin{figure}
   \centering
    \includegraphics[width=9.0cm]{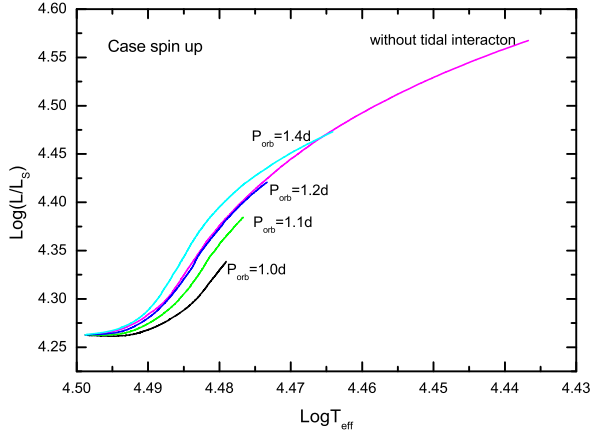}
      \caption{Evolutionary tracks for 15 M$_\odot$ models with (spin up case) and without tidal interaction. Curves for different initial orbital periods are shown. For close binaries, only the part of the track corresponding to the period before the Roche Lobe Overflow is plotted.}
         \label{hr2}
   \end{figure}

The impact on the evolutionary tracks can be seen in Fig.~\ref{hr2}.
There is a shift of tidally accelerated models towards redder positions compared to the model without tidal interaction, for periods shorter than about 1.2-1.3 days. It comes from the fact that globally the stars is rapidly spun up and the centrifugal acceleration becomes non negligible.
 The centrifugal acceleration balances the gravity. This tends to make the star to follow an evolution corresponding to
a lower initial mass, thus with a lower luminosity and being redder.
For the period 1.4 days, one sees that the track is only very slightly overluminous with respect to the track without tidal interaction. In that case, the hydrostatic effects and those of chemical mixing
are more or less compensating each other. This produces a track not far from the one computed without tidal interaction.

\section{Conclusions and observable consequences}

From the above results, we can deduce some interesting general trends and consequences :

1) At least up to the RLOF, the direction of the meridional circulation is always in a sense that counteracts the effects of the tidal interaction.
Meridional currents are spinning up the surface when it is braked down (spin down case), and are braking it down when it is accelerated (spin up case).
This illustrates how circulation is driven by what happens at the surface, tending to restore previous equilibrium. These trends seem to be a general property of meridional circulation.

2) For correctly modeling the effects of tidal interactions, a proper account of the transport of angular momentum by advection through meridional currents
is essential.

3) {\bf The effective time for obtaining $\Omega=\omega_{\rm orb}$
may extend over periods covering a significant fraction of the
Main-Sequence lifetime.}

4) Whatever tides brake down the surface or spin it up, they boost the shear diffusion coefficient and the surface nitrogen enrichments.

5) How these effects vary as a function of the initial period depends on whether tidal interactions brake down or spin up the star. In case of braking down, the effects  obtained at the end of synchronisation on the surface enrichments are enlarged when the initial period increases. The contrary occurs in case of tidal spin up.

From the point of view of possible interesting observable consequences of tidal interactions, we can note :

1) {\bf Spin-up systems studied here encounters RLOF before being
synchronized.}

2) The possibility through asteroseismology to probe the internal rotation rate of stars during their synchronization period or after synchronization (but before
any RLOF) would provide extremely interesting constraints on the present models.

3) Tidal braking produces stars which are strongly nitrogen rich, slowly rotating and presenting a high surface gravity.
Such a braking can be invoked to explain some slow, non evolved rotators  with strong nitrogen enrichment.

4) Tidal spinning up may produce fast rotators before the RLOF episode.
After RLOF,  the loss of the envelope will probably considerably  slow down the star.
Thus it will be likely not possible to produce fast rotators following a homogeneous evolution
through tidal interaction. The process can be initiated by tidal spin up but, at a given point, RLOF will stop it.

5) In tidal spin up, part of the angular momentum acquired by the primary will be locked into the core due to the redistribution
of the angular momentum. Is there any link with Gamma Ray Bursts? This has to be further studied in forthcoming works.

Let us note that the results would be very different in case some process would force the star to rotate as a solid body
in its interior (this might be due for instance to a strong internal magnetic field). We can suspect that the results would be
qualitatively similar to those obtained when a magnetic braking is applied at the surface of a solid body rotating star (Meynet et al. 2011):
a very fast change of the surface velocity (short time for obtaining $\Omega=\omega$) and no tidal mixing during the synchronisation period.
This will be studied in a forthcoming paper.

\begin{acknowledgements}
This work was sponsored by Key Laboratory for the Structure and
Evolution of Celestial Objects, Chinese Academy of Sciences,(Grant
No. OP201107), the Key Foundation at Guizhou education department
(No 2010002). We thank Dr. Patrick Eggenberger for having made
available his routine for the change of angular momentum in the
outer layers and for very useful remarks helping to improve this
paper. \textbf{We are very grateful to professor Jean-Paul Zahn
(referee) for his valuable suggestions and insightful remarks, which
have improved this paper greatly. Also we thank the editor for
his/her professional and excellent work.}
\end{acknowledgements}

\end{document}